\documentclass[aps,amsfonts,nofootinbib]{revtex4}
\usepackage{epsfig}
\usepackage{graphicx}
\usepackage{amsmath}
\usepackage{amsbsy}
\usepackage{color}
\usepackage{epsfig}
\usepackage{graphicx}
\usepackage{amsmath}
\usepackage{amssymb}
\usepackage{amsbsy}
\newcommand{\ee}{\end{equation}}
\newcommand{\bb}{\begin{equation}}
\newcommand{\eqb}{\begin{eqnarray}}
\newcommand{\eqf}{\end{eqnarray}}

\def\p{\mathbf{p}}

\newcommand{\1}{{\'{\i}}}

\def\1{\'{\i}}

\begin{document}
\title{Noncommutativity in (2+1)-dimensions and the Lorentz group}

\author{H.\ Falomir}
\email{falomir@fisica.unlp.edu.ar, }
\affiliation{IFLP - CONICET and Departamento de  F\'{\i}sica, Facultad de Ciencias Exactas de la UNLP, \\
C.C.\ 67, (1900) La Plata, Argentina}
\author{J.\ Gamboa}
\email{jgamboa55@gmail.com}
\affiliation{Departamento de  F\'{\i}sica, Universidad de  Santiago de
  Chile, Casilla 307, Santiago, Chile}
\author{M.\ Loewe}
\email{mloewe@fis.puc.cl}
\affiliation{Facultad    de   F\'{\i}sica,    Pontificia   Universidad
  Cat\'olica de Chile, Casilla 306, Santiago 22, Chile\\ and \\
  Centre for Theoretical Physics and Mathematical Physics, University of Cape Town, Rondebosch 7700, South Africa}
\author{F.\ M\'{e}ndez}
\email{fernando.mendez.f@gmail.com}
\affiliation{Departamento de  F\'{\i}sica, Universidad de  Santiago de
  Chile, Casilla 307, Santiago, Chile}
\author{F.\ Vega}
\email{federicogaspar@gmail.com}
\affiliation{IFLP - CONICET and Departamento de  F\'{\i}sica, Facultad de Ciencias Exactas de la UNLP, \\
C.C.\ 67, (1900) La Plata, Argentina}

\begin{abstract}

In this article we considered models of particles living in a three-dimensional space-time with a nonstandard noncommutativity induced by shifting canonical coordinates and momenta with generators of a unitary irreducible representation of the Lorentz group. The Hilbert space gets the structure of a direct product with the representation space, where we are able to construct operators which realize the algebra of Lorentz transformations.

We study the modified Landau problem for both Schr\"{o}dinger and Dirac particles, whose Hamiltonians are obtained through a kind of non-Abelian Bopp's shift of the dynamical variables from the ones of the usual problem in the normal space. The spectrum of these models are considered in perturbation theory, both for small and large noncommutativity parameters.

We find no constraint between the parameters referring to no-commutativity in coordinates and momenta but they rather play similar roles.

Since the representation space of the unitary irreducible representations  $SL(2,\mathbb{R})$ can be realized in terms of spaces of square-integrable functions, we conclude that these models are equivalent to quantum mechanical models of particles living in a space with an additional compact dimension.

\vskip 0.3cm


\end{abstract}
%\pacs{PACS numbers: 03.65.-w; 11.30.Cp; 02.40.Gh}
\date{\today}
%\date{August 31, 2012}
\maketitle

\section{Introduction}

Space-time non-commutativity is an old idea \cite{Snyder,Yang} which has been revived in recent years within the context of string theory \cite{Douglas} and has attracted much attention in diverse areas such as Mathematics \cite{Connes1,Connes2}, Theoretical Physics  \cite{Witten,Seiberg}, Phenomenology \cite{Hinchliffe} or Condensed Matter \cite{Dayi,Bellissard}.

In the conventional version of noncommutative (NC) space-time, the coordinate operators satisfy the algebra
\begin{equation}\label{cha1}
    \left[ x^\mu , x^\nu \right] = \imath \theta^{\mu \nu}\,,
\end{equation}
where $\theta^{\mu \nu}$ is a real constant antisymmetric matrix, not a tensor. This is the situation realized in string theory in the presence of a background antisymmetric tensor field \cite{Seiberg}. But, clearly, such $\theta^{\mu \nu}$ define preferred directions in a given Lorentz frame and, thus, the assumption in Eq.\ (\ref{cha1}) produces a violation of Lorentz invariance \cite{Yang,Green,Polchinski,Szabo,muchos}.

In \cite{Carlson}, a different class of noncommutative theories have been considered in order to recover Lorentz invariance. In these models, the parameter $\theta^{\mu \nu}$ in the right hand side of Eq.\ (\ref{cha1}) is promoted to an operator that transforms as a Lorentz tensor. This algebra can be interpreted as a contraction of the Lorentz-invariant algebra due to Snyder \cite{Snyder}, taking $\theta^{\mu \nu}$ as proportional to the generators of the Lorentz group.

These ideas have been put in practice in a series of papers \cite{otrosLI}, employing in particular the Doplicher, Fredenhagen and Roberts algebra \cite{Doplicher}, in which the operators $\theta^{\mu \nu}$ are considered as part of the ordinary coordinates of an enlarged ten-dimensional space-time, with the assumptions that the triple commutator of coordinate operators $x^\mu$ vanishes. This algebra was later extended by Amorim \cite{Amorin} through the introduction of momenta canonically conjugate to these new coordinates.

Noncommutative Quantum Mechanics (NCQM) \cite{NCQM} is a simple scenario to explore the properties of NC spaces.   In addition to the non-commutativity of the position operators previously referred to and the study of representations of the algebra of the NC space-time coordinates \cite{Doplicher,newton}, non-commutativity in the momenta commutators algebra have also been considered, in relation with the deformation quantization of Poissonian structures \cite{Rie} and as a kind of magnetic quantization \cite{Ift,Man}.

These researches have stimulated the construction of new models in quantum mechanics \cite{NCQM}, allowing to explore new ideas in several situations of interest. For example, some models based on a kind of nonstandard deformation of the Heisenberg algebra, which can be realized by shifting the dynamical variables with spin variables, have been studied in \cite{SpinNC1,SpinNC2}. A similar deformation in the commutators among momenta can be interpreted as the introduction of constant non-Abelian magnetic fields \cite{weiss}.

This kind of noncommutativity in the phase space, where the number of degrees of freedom is enlarged by shifting the usual dynamical variables by adding them a generator in the Lie algebra of a non-Abelian group \cite{SpinNC1,SpinNC2}, have been employed in the formulation of some interesting quantum-mechanical models. For example, these ideas have even found application in the description of graphene, a new material recently experimentally obtained which behaves as a two-dimensional system. In \cite{graphene} it has been studied a two-dimensional continuous model which takes some elements of the tight-binding model for this material and reproduces the anomalous integer quantum Hall effect characteristic of graphene.

\smallskip

In the following we consider a model with this kind of nonstandard noncommutativity, where both coordinates and momenta get deformed commutators. These deformations are taken as proportional to the generators of the Lorentz group in some irreducible representation (\emph{irrep}) rather than proportional to the generators of this group on the space-time themselves, as in \cite{Carlson}. As we will see, under these conditions \textit{total} generators of the Lorentz transformations can be constructed which correctly transform all the operators, giving thus a realization of the Lie algebra of this group on the Hilbert space of the quantum-mechanical system.

These deformations of the Heisenberg algebra can be effectively realized in a NC three-dimensional space-time by simply shifting the ordinary (commutative) dynamical variables by terms proportional to the generators in an \emph{irrep} of the Lorentz group on the 2+1-Minkowski space, isomorphic to $SL(2,\mathbb{R})/\mathbb{Z}$, since in this case the dimension of the group coincides with the dimension of the space-time. A similar idea has recently been developed in a four-dimensional NC space through shifts in the coordinates proportional to the Pauli-Lubanski vector \cite{Ferrari}.

Let us mention that three-dimensional theories have regained interest in recent years since it was found that this is the only non-critical dimension where strings can be quantized consistently \cite{MT1,MT2,MT3,MT4}. Indeed, it was shown  that light-cone gauge quantization in the three-dimensional space preserves Lorentz invariance without the need for any longitudinal modes \cite{MT1}. It was also found that some states necessarily have irrational spin, {\it i.e}.\ the spectrum contains anyons. This led to the consideration of infinite-dimensional irreducible representations of the universal cover of the Lorentz group on the three-dimensional Minkowski space \cite{MT3}.

\medskip

In this  article we consider the behavior of Schr\"{o}dinger and Dirac particles in a NC 2+1-dimensional space, in which the \emph{phase-space noncommutativity} is induced by shifting the usual dynamical variables by generators in an irreducible representation of the (noncompact) Lie group $SL(2,\mathbb{R})$ \cite{Bargmann,Binegar,Jackiw}. This means that the Hilbert space has the structure of a direct product, where one factor corresponds to the component of the state vectors in the representation space of the \emph{irrep} considered.

Notice that if we demand the noncommuting phase space variables to be Hermitian operators, we are constrained to consider unitary \emph{irrep}'s of this group, which are not of finite dimension. Moreover, since the representation space of the unitary \emph{irrep}'s of $SL(2,\mathbb{R})$ can be explicitly realized in terms of spaces of functions defined on the unit circle or analytic functions on the unit open disk, as discussed in Appendix \ref{irrep-unit} (See reference \cite{Bargmann}), the models to be considered turn out to be equivalent to quantum mechanical systems living in a space with an additional (compact) dimension.

\smallskip

The structure of the paper is as follows. In Section \ref{NCspace} we set up the noncommutativity properties of the three-dimensional space, realize the deformed Heisenberg algebra though a kind of non-Abelian Bopp's shift in the dynamical variables and discuss the construction of the generators of the Lorentz group on the Hilbert space. In Section \ref{Schrodinger} we consider the Hamiltonian of a Schr\"{o}dinger particle in the presence of an external ($U(1)$) magnetic field, both in the normal and the NC space. We explain the characteristics of the spectrum in both the small and large NC parameters limits and discuss the relation with a system in a space-time with an additional dimension. In Section \ref{Dirac} we make the same analysis for the Hamiltonian of a Dirac particle. In Section \ref{conclusion} we state our conclusions and, for completeness, in Appendix \ref{SL2R} we briefly review the Lorentz group in 2+1-dimensions and the unitary \emph{irrep}'s  of $SL(2,\mathbb{R})$.

\section{Setting of the problem}\label{NCspace}

According to the ideas previously exposed, we consider the modified Heisenberg algebra of the (Hermitian) dynamical variables given by
\begin{equation}
\label{d30}
\begin{array}{lll} \displaystyle
\left[{\hat  x}_\mu,{\hat x}_\nu \right]  =- i\theta^2  \epsilon_{\mu \nu \rho} s^\rho \,,  & \quad  &   \left[{\hat   p}_\mu,{\hat   p}_\nu\right]   =   -i   \kappa^2
\epsilon_{\mu \nu \rho} s^\rho\,,
\\  \\ \displaystyle
\left[{\hat  x}_\mu,{\hat p}_\nu\right]  = i(\eta_{\mu \nu}  -  \kappa \theta
\epsilon_{\mu \nu \rho} s^\rho)\,, & \quad &
\left[ {\hat  x}_\mu, s_\nu \right] = - i \theta\epsilon_{\mu \nu \rho} s^\rho
\,,
\\  \\ \displaystyle
\left[  {\hat  p}_\mu,   s_\nu  \right]  =   -i  \kappa \epsilon_{\mu \nu \rho} s^\rho
\,, & \quad  &
\left[s_\mu,s_\nu \right] =- i\epsilon_{\mu \nu \rho} s^\rho\,,
\end{array}
\end{equation}
where $s_\mu\,, \mu=0,1,2$ are the generators of an \emph{irrep} of $SL(2,\mathbb{R})$ (See Appendix \ref{SL2R}) and $\theta$ and $\kappa$ play the role of ultraviolet and infrared scales respectively.

This deformation can be realized through a kind of \emph{non-Abelian Bopp's shift} given by
\begin{equation}\label{shift}
    \hat{x}_\mu\rightarrow x_\mu+\theta s_\mu\,,\quad \hat{p}_\mu\rightarrow p_\mu + \kappa s_\mu\,,
\end{equation}
in terms of dynamical variables satisfying the usual Heisenberg algebra and the generators of a unitary \emph{irrep} of $SL(2,\mathbb{R})$, which satisfy the commutation relations
\begin{equation}\label{31}
   \begin{array}{ll} \displaystyle
        \left[ x^\mu , x^\nu \right]=0\,, \quad    \left[ x^\mu , p_\nu \right]=\imath \delta_\nu^\mu\,,
  \\  \\ \displaystyle
  \left[ p_\mu , p_\nu \right]=0 \,, \quad
  \left[ x^\mu , s_\nu \right]=0\,,
  \\  \\ \displaystyle
   \left[ p_\mu , s_\nu \right]=0 \,, \quad
   \left[ s_\mu , s_\nu \right]= - \imath \epsilon_{\mu \nu \lambda} s^\lambda\,.
   \end{array}
\end{equation}
Here, $a_\lambda= \eta_{\lambda\gamma}a^\gamma$, with $\eta:=$diag$(1,-1,-1)$, the metric tensor in the 2+1-dimensional Minkowski space.

Notice that we must consider a \emph{unitary} representation of $SL(2,\mathbb{R})$ (non Abelian Lie group isomorphic to $SU(1,1)$), in order to have Hermitian operators $\hat{x}^\mu$ and $\hat{p}_\mu$ representing coordinates and momenta in the noncommutative phase space. Since this is a noncompact Lie group, its irreducible unitary representations are not of finite dimension. The irreducible representations of $SL(2,\mathbb{R})$ are briefly reviewed in Appendix \ref{SL2R}.

\medskip

If we define $L_{\mu\nu}:= x_\nu p_\mu - x_\mu p_\nu$ and $L^\mu:=\frac{1}{2}\, \epsilon^{\mu \nu \lambda}L_{\nu\lambda}$, we get
\begin{equation}\label{32}
  \begin{array}{ll}\displaystyle
\left[ L_\mu , x_\nu \right]= -\imath \epsilon_{\mu \nu \lambda} x^\lambda\,, \quad
\left[ L_\mu , p_\nu \right]= -\imath \epsilon_{\mu \nu \lambda} p^\lambda\,,
\\ \\ \displaystyle
\left[ L_\mu , L_\nu \right]= -\imath \epsilon_{\mu \nu \lambda} L^\lambda\,, \quad
\left[ L_\mu , s_\nu \right]=0\,.
\end{array}
\end{equation}

Since the operators $L_\mu$ satisfy the same commutation relations as the $s_\mu$, we can define new  generators of $sl(2,\mathbb{R})$, $M_\mu:=L_\mu+s_\mu$, for which we get (See Eq.\ (\ref{111}))
\begin{equation}\label{33}
   \begin{array}{c}\displaystyle
\left[ M_\mu , x_\nu \right]= -\imath \epsilon_{\mu \nu \lambda} x^\lambda\,,\quad
\left[ M_\mu , p_\nu \right]=- \imath \epsilon_{\mu \nu \lambda} p^\lambda\,,
\\ \\ \displaystyle
\left[ M_\mu , s_\nu \right]= -\imath \epsilon_{\mu \nu \lambda} s^\lambda\,,\quad
\left[ M_\mu , M_\nu \right]= -\imath \epsilon_{\mu \nu \lambda} M^\lambda\,.
\end{array}
\end{equation}

Then, these operators are a realization on the Hilbert space of the system of the generators of the Lorentz transformations in the 2+1-dimensional noncommutative space we are considering. Indeed, it is easy to see that
\begin{equation}\label{333}
   \begin{array}{c}\displaystyle
\left[ M_\mu , \hat{x}_\nu \right]= -\imath \epsilon_{\mu \nu \lambda} \hat{x}^\lambda\,,\quad
\left[ M_\mu , \hat{p}_\nu \right]=- \imath \epsilon_{\mu \nu \lambda} \hat{p}^\lambda\,.
%\\ \\ \displaystyle
%\left[ M_\mu , {s}_\nu \right]= -\imath \epsilon_{\mu \nu \lambda} {s}^\lambda\,.
\end{array}
\end{equation}

\medskip

Now, our strategy to formulate a model in the noncommutative space will be, given a Hamiltonian $H(\mathbf{p},\mathbf{x})$ in the usual (commutative) Minkowski space, to generalize it by taking $H(\mathbf{\hat{p}},\mathbf{\hat{x}})$. This problem will then be analyzed through the replacements in Eq.\ (\ref{shift}).

Notice that the second commutator in Eq.\ (\ref{d30}) for the two spatial momenta, which maps onto $\left[p_1+\kappa s_1 , p_2+\kappa s_2\right]   =   - \imath   \kappa^2   s_0$  where $s_0$ is the generator of rotations on the (spatial) plane in the \emph{irrep} considered, can equivalently be interpreted as a consequence of the presence of a constant and uniform non-Abelian magnetic field. Therefore, the models to be studied in the following are a kind of generalizations of the non-Abelian Landau problem \cite{weiss}. This point of view was employed in \cite{graphene} to construct a continuous model which incorporates next-to-leading contributions from the dispersion relation of the tight-binding model for graphene. In the present approach we also incorporate noncommutativity in the temporal momentum component.

On the other hand, a commutation relation like the first one in Eq.\ (\ref{d30}) among coordinate operators in the three-dimensional Euclidean space, but with generators of $SU(2)$ in the right hand side, allows to simulate dipolar interactions and lead to models with infinitely degenerate ground state and spontaneous symmetry breaking \cite{SpinNC1} and find application, for example, in the description of triplet superconductivity  \cite{triplet}. Contrary to that case, in the present article we need to consider infinite-dimensional unitary \emph{irrep}'s of a noncompact group.

\medskip

As previously mentioned, in the following we construct models of Schr\"{o}dinger and Dirac particles in the NC extension of the  2+1-dimensional Minkowski phase space described by Eqs.\ (\ref{d30}), and study the implications this generalization may have. In particular, we will get a space of state vectors which is the direct product of the Hilbert space for the systems in the usual commutative space with the representation space of a unitary \emph{irrep} of $SL(2,\mathbb{R})$. Since these representation spaces can be realized as spaces of square-integrable functions (functions on the unit circle or analytic functions on the open unit disk according to the particular \emph{irrep} considered, as discussed in Appendix \ref{irrep-unit}), these models can also be interpreted as describing particles living in spaces with an additional (compact) spatial dimension.

\section{Schr\"{o}dinger particles}\label{Schrodinger}

To establish our notation, let us first consider the Schr\"{o}dinger Hamiltonian for the Landau problem. We take the external electromagnetic field as given by $A_0=0$ and $A_i(\mathbf{x})=-\frac{B}{2} \epsilon_{i j} x^j$.
Then, we have for a particle of mass $M$
\begin{equation}\label{S1}
    2 M H:=\left( p_i - e A_i(x) \right)^2=\left( p_i - \frac{eB}{2} \epsilon_{i j} x_j\right)^2 \,,
\end{equation}
operator which commutes with the generator of rotations on the plane, $L_0=L_{12}$ (See Ec.\ (\ref{32})).

As is known, this operator can be given the form of the Hamiltonian of a harmonic oscillator through the canonical transformation of variables
\begin{equation}\label{S2}
    q:=\frac{p_1}{\sqrt{{eB}}}+\frac{\sqrt{eB}}{2} x^2\,, \quad p:=\frac{p_2}{\sqrt{{eB}}}-\frac{\sqrt{eB}}{2} x^1\,,
\end{equation}
for which we have
\begin{equation}\label{S3}
    \left[ q,p \right]=  \left[ \frac{p_1}{\sqrt{{eB}}}+\frac{\sqrt{eB}}{2} x^2 , \frac{p_2}{\sqrt{{eB}}}-\frac{\sqrt{eB}}{2} x^1 \right] = \imath\,.
\end{equation}
Indeed, we get
\begin{equation}\label{S4}
     H =\left( \frac{eB}{M} \right) \frac{\left( p^2 + q^2 \right)}{2}= \left(\frac{ eB}{M} \right) \left[ a^\dagger a + \frac{1}{2} \right]
\end{equation}
where, as usual,
\begin{equation}\label{S5}
    \quad a= \frac{q+\imath p}{\sqrt{2}}\,, \quad a^\dagger= \frac{q-\imath p}{\sqrt{2}}\,, \quad \left[ a , a^\dagger \right]=1\,.
\end{equation}

We also introduce the independent set of canonical variables
\begin{equation}\label{PQ1}
    Q:=\frac{p_2}{\sqrt{{eB}}}+\frac{\sqrt{eB}}{2} x^1\,, \quad P:=\frac{p_1}{\sqrt{{eB}}}-\frac{\sqrt{eB}}{2} x^2\,,
\end{equation}
which satisfy
\begin{equation}\label{PQ2}
    \begin{array}{c} \displaystyle
       \left[P , Q \right]=-\imath\,, \quad
       \left[P , p \right]=0\,, \quad
       \left[P , q \right]=0\,,
       \\ \\ \displaystyle
      \left[Q , p \right]=0\,, \quad
       \left[Q , q \right]=0\,.
    \end{array}
\end{equation}

Similarly, we define creation and destruction operators as
\begin{equation}\label{S555}
    \begin{array}{c} \displaystyle
      \quad b= \frac{Q+\imath P}{\sqrt{2}}\,, \quad b^\dagger= \frac{Q-\imath P}{\sqrt{2}}\,,
    \end{array}
\end{equation}
which satisfy
\begin{equation}\label{S55555}
    \begin{array}{c} \displaystyle
    \left[ b , b^\dagger \right]=1\,, \quad
    \left[ b , a \right]= 0  \,, \quad \left[ b , a^\dagger \right] = 0\,.
    \end{array}
\end{equation}

The Hamiltonian eigenvectors are then given by
\begin{equation}\label{S6}
    \left| n,n_b  \right\rangle = \frac{\left(a^\dagger\right)^n}{\sqrt{n!}}
    \frac{\left(b^\dagger\right)^{n_b} }{\sqrt{n_b !}}\left| 0,0 \right\rangle \,,
\end{equation}
with $n,n_b =0,1,2,\cdots$, and the corresponding eigenvalues by
\begin{equation}\label{S6666}
    \quad E_n=\left( \frac{eB}{M} \right) \left[ n + \frac{1}{2} \right]\,,
\end{equation}
degenerate in the index $n_b $. Here, $\left| 0 , 0\right\rangle \leftrightarrow
\psi_0(q,Q)=\pi^{-\frac{1}{2}}{e^{-\frac{q^2+Q^2}{2}}}$ and $\left\langle n', n_b ' |
n , n_b  \right\rangle=\delta_{n',n} \delta_{n_b',n_b}$.

For $L_0$ we get
\begin{equation}\label{PQ3}
    L_0=x^1 p_2 - x^2 p_1 = -\frac{1}{2}\left( p^2 + q^2 \right)+\frac{1}{2}\left( P^2 + Q^2 \right)=b^\dagger b-a^\dagger a\,,
\end{equation}
and for its eigenvalues the difference $l=n_b -n\in \mathbb{Z}$.

\subsection{Extension to the noncommutative space}

The generalization of this system to the noncommutative space defined by Eq.\ (\ref{d30}) is
\begin{equation}\label{S7}
     2 M \hat{H}:=\left( \hat{p}_i - \frac{eB}{2} \epsilon_{i j} \hat{x}_j\right)^2 +2M\kappa s_0
     = \left[ p_i+\kappa s_i - \frac{eB}{2} \epsilon_{i j} \left(x_j +\theta s_j\right) \right]^2 +2M\kappa s_0
\end{equation}
(where the last term in the right hand side comes from the shift applied to $p_0$ in Eq.\ (\ref{shift})), which commutes with $M_0=L_0+s_0$ as follows from Ec.\ (\ref{333}).

If we define $\hat{\pi}_i:=\hat{p}_i - \frac{eB}{2} \epsilon_{i j} \hat{x}_j$, we get from (\ref{d30})
\begin{equation}\label{Spi}
    \left[ \hat{\pi}_1 , \hat{\pi}_2 \right]=-\imath e B -\imath \left[\kappa^2+\left(\theta \frac{eB}{2}\right)^2 \right] s_0\,,
\end{equation}
which shows that the model  we are considering can also be interpreted as the introduction, besides the $U(1)$ magnetic field, of another
constant nonabelian \emph{magnetic field} in the \emph{time direction} (spatial rotations) of the Lie algebra $sl(2,\mathbb{R})$.

\medskip

In terms of $a^\dagger$ and $a$ in Eq.\ (\ref{S5}) and the Hermitian generators $s_\mu$ in Eq.\ (\ref{31}), this Hamiltonian reads as
\begin{equation}\label{S8}
    \begin{array}{c} \displaystyle
      2 M \hat{H}
     =2MH+2M\kappa s_0
     +\sqrt{2eB}\left\{\left[ \kappa+\imath \theta \frac{eB}{2} \right] a^\dagger s_+ +
     \left[ \kappa-\imath \theta \frac{eB}{2} \right] a s_- \right\}+
     \\ \\ \displaystyle
     +\left[\kappa^2+\left(\theta \frac{eB}{2}\right)^2 \right]\left( {s_0}^2-{\mathbf{s}}^2\right) \,,
    \end{array}
\end{equation}
where
\begin{equation}\label{gen-sl2r}
    s_\pm:=s_1\pm \imath s_2\,, \quad \mathbf{s}^2:={s_0}^2-{s_1}^2-{s_2}^2\,.
\end{equation}

Notice that the parameter $\kappa$ appears as an energy scale for the \emph{internal} degrees of freedom while the pure number $ \frac{\theta eB}{2\kappa}$ (for $\kappa\neq 0$) is a measure of the relative strength of noncommutativity in coordinates and momenta with respect to the applied external magnetic field.

\medskip

In Appendix \ref{irrep-unit} we give a brief review of the unitary irreducible representations of $sl(2,\mathbb{R})$. For a given unitary \emph{irrep}, the representation space is subtended by the basis of eigenvectors of $s_0$ and
$\mathbf{s}^2$,
\begin{equation}\label{16666}
    {\mathbf{s}^2} \left|\lambda,m\right\rangle = \lambda \left|\lambda,m\right\rangle\,, \quad
    s_0 \left|\lambda,m\right\rangle = m \left|\lambda,m\right\rangle\,,
\end{equation}
where $\lambda$ and $m$ are real numbers.

The Hilbert space is then the linear span of the vectors of the form
\begin{equation}\label{S9}
    \left| n,n_b ; \lambda,m \right\rangle:=\left|n,n_b \right\rangle\otimes\left| \lambda, m \right\rangle \,,
\end{equation}
which are simultaneously eigenvectors of $H$, $L_0$, $\mathbf{s}^2$ and $s_0$, normalized so as to satisfy
\begin{equation}\label{S10}
   \left\langle n,n_b ; \lambda,m |  n',n_b '; \lambda,m' \right\rangle = \delta_{n,n'}\delta_{n_b ,n_b '}\delta_{m,m'}\,.
\end{equation}

Let us recall that
\begin{equation}\label{S11}
    a^\dagger \left|n,n_b \right\rangle = \sqrt{n+1} \left|n+1,n_b \right\rangle\,, \quad a \left|n,n_b \right\rangle = \sqrt{n} \left|n-1,n_b \right\rangle\,,
\end{equation}
and (See Eq.\ (\ref{20}))
\begin{equation}\label{S12}
    s_\pm \left| \lambda, m \right\rangle = \sqrt{m(m\pm 1)-\lambda}\,   \left| \lambda, m\pm 1 \right\rangle\,.
\end{equation}

We also have that $\left[ \hat{H} , M_0 \right]=0$, where $M_0=L_0+s_0$ has eigenvalues $j=l+m=n_b -n+m$, integer or half-integer according to the \emph{irrep} of $SL(2,\mathbb{R})$ considered \cite{Bargmann}.
Indeed,  let us call $z:= \kappa+\imath \theta \frac{eB}{2} $; then, it is straightforward to get
\begin{equation}\label{M0-1}
   \begin{array}{c} \displaystyle
      \left[2M \hat{H} , L_0 \right]
     =\sqrt{2 e B} \left[ \left( z a^\dagger s_+ +
      \bar{z} a s_- \right) , - a^\dagger a  \right] = \sqrt{2 e B} \left\{ z a^\dagger s_+ - \bar{z} a s_-  \right\}\,,
   \end{array}
\end{equation}
and
\begin{equation}\label{M0-2}
   \begin{array}{c} \displaystyle
      \left[2M \hat{H} , s_0 \right]
     =\sqrt{2 e B} \left[ \left( z a^\dagger s_+ +
      \bar{z} a s_- \right) , s_0  \right] = \sqrt{2 e B} \left\{ -z a^\dagger s_+ + \bar{z} a s_-  \right\}\,.
   \end{array}
\end{equation}

Moreover, $\left[ \hat{H} , b^\dagger b\right]=0$. Then, for given values of $\lambda$, $j$ and $n_b $, we can give the following development for the $\hat{H}$'s eigenvectors,
\begin{equation}\label{S13}
    \left| \psi_{E,j,n_b }\right\rangle = \sum_{n-m=n_b -j} C_{n,m}   \left| n,n_b ; \lambda,m \right\rangle\,.
\end{equation}

\medskip

From Eq.\ (\ref{S8}), one straightforwardly gets the recursion relation for the coefficients
\begin{equation}\label{S14}
  \begin{array}{c} \displaystyle
     \left\langle n,n_b ; \lambda,m  \right| 2M (\hat{H} -E)  \left| \psi_{E,j,n_b }\right\rangle=
     \left\{ 2eB(n+1/2)-2M(E-\kappa m)+\bar{z} z \left( m^2-\lambda\right)  \right\} C_{n,m}+
     \\ \\ \displaystyle
     +z \sqrt{2eB} \sqrt{n}\sqrt{m(m-1)-\lambda}\,  C_{n-1,m-1} +
      \bar{z}  \sqrt{2eB}\sqrt{n+1}\sqrt{(m+1)m-\lambda}\,  C_{n+1,m+1}=0\,,
  \end{array}
\end{equation}
where $m=j-n_b +n$.

Notice that, for $z=0$, this recurrence gives immediately the usual Landau levels,
\begin{equation}\label{chequeo}
    C_{n,m} \left[e B (n+1/2)-M E \right]=0 \quad \Rightarrow \quad E= \frac{e B}{M} \left( n + \frac{1}{2}\right)\,.
\end{equation}

\medskip

It is not evident how to solve the recurrence in Eq.\ (\ref{S14}) in the general case. This problem simplifies, for example, for the case of a unitary \emph{irrep} in the discrete series (See Appendix \ref{discretas}), with $\lambda=k(k-1)$ and $m \leq -k$, where $k$ is a positive integer or half-integer, since in this case the right hand side of Eq.\ (\ref{S13}) reduces to a finite sum. Indeed,
\begin{equation}\label{S144}
    m=j-n_b +n\leq-k \quad \Rightarrow \quad 0\leq n \leq  n_b -j-k\,,
\end{equation}
which requires that $j-n_b \leq -k$ in order to have a nontrivial solution.

In this case we have
\begin{equation}\label{S15}
    \left| \psi_{E,j,n_b }\right\rangle = \sum_{n=0}^{n_b -k-j} C_{n,j-n_b +n}   \left| n,n_b ; k(k-1),j-n_b +n \right\rangle\,,
\end{equation}
and the eigenvalues problem reduces to a matricial one. Notice that the eigenvalues depend on $j$ and $n_b $ only through the difference $J:=j-n_b $, which gives rise to the infinite degeneracy characteristic of the Landau problem.

On the contrary, for an \emph{irrep} of the discrete series with $m=j-n_b +n\geq k$, then $n\geq k-j+n_b $ and one must determine a whole series.

\medskip

If, for example, we take $j-n_b =-1/2$ and $m\leq -k$ for the \emph{irrep} with $k=1/2$, we have a unique nontrivial solution with $n=0$ and $m=-\frac{1}{2}$, % para $l=j-m=n_b $,
\begin{equation}\label{S16}
     \left| \psi_{E_0,n_b -\frac{1}{2},n_b }\right\rangle =C_{0,-\frac{1}{2}}  \left| 0,n_b ; -\frac{1}{4},-\frac{1}{2} \right\rangle\,,\quad
     {\rm with}\quad E_0=\frac{eB}{M}\left(\frac{1}{2}\right) - \frac{\kappa}{2}- \frac{\bar{z} z}{4M}\,,
\end{equation}
with an infinite degeneracy in the index $n_b =0,1,2,\dots$ In this case, the noncommutativity of the phase space produces a negative shift in the energy of the fundamental Landau level.

For the same \emph{irrep} and with $j-n_b =-\frac{3}{2}$, the solutions belong to a two-dimensional subspace (for each $n_b $) containing the independent eigenvectors
\begin{equation}\label{S177}
    \begin{array}{c}
    \left|\psi_{E_1,n_b -\frac{3}{2},n_b }\right\rangle = \left( 1+\frac{M\kappa}{eB}+O(z^2) \right) \left| 0,n_b ; -\frac{1}{4},-\frac{3}{2} \right\rangle
      - \left\{\frac{z }{\sqrt{2eB}}+O\left(z^2\right)\right\}
      \left| 1,n_b ; -\frac{1}{4},-\frac{1}{2} \right\rangle
   \\  \\
     \left|\psi_{E'_1,n_b -\frac{3}{2},n_b }\right\rangle = \left\{\frac{\bar{z} }{\sqrt{2eB}} +
     O\left(z^2\right)\right\}\left| 0,n_b ; -\frac{1}{4},-\frac{3}{2} \right\rangle
      +\left(1+O(z^2)\right) \left| 1,n_b ; -\frac{1}{4},-\frac{1}{2} \right\rangle
    \end{array}
\end{equation}
corresponding to eigenvalues which are more involved functions of the noncommutativity parameters and up to quadratic order in $|z|$ reduce to
\begin{equation}\label{S18}
    \begin{array}{c} \displaystyle
      E_1=\frac{B e}{ M}\left(\frac{1}{2}\right)-\frac{3\kappa}{2}+ \frac{3  \bar{z} z }{4 M} +O\left(z ^3\right)\,,
     \\ \\ \displaystyle
      E_1'=\frac{ B e}{ M}\left(1+\frac{1}{2}\right)-\frac{\kappa}{2}+ \frac{3 \bar{z}z}{4 M} +O\left(z ^3\right)\,,
    \end{array}
\end{equation}
again degenerated in the index $n_b $. Then, here we also find an $O(z^2)$ shift from the first Landau levels in the normal commutative plane.

Similarly, for $j-n_b =-\frac{5}{2}$ the eigenvalues (degenerate in $n_b $) up to quadratic order in $|z|$ are
\begin{equation}\label{S19}
    \begin{array}{c} \displaystyle
      E_2= \frac{B e}{ M}\left(\frac{1}{2}\right)-\frac{5\kappa}{2}+\frac{5 \bar{z}z}{4M}+O\left(z ^3\right) \,,
     \\ \\ \displaystyle
      E_2'= \frac{ B e}{ M}\left(1+\frac{1}{2}\right)-\frac{3\kappa}{2}+\frac{9 \bar{z}z}{4 M}+O\left(z ^3\right) \,,
         \\ \\ \displaystyle
         E_2''=  \frac{B e}{ M}\left(2+\frac{1}{2}\right)-\frac{\kappa}{2}+\frac{5 \bar{z}z}{4 M}+O\left(z ^3\right)\,.
    \end{array}
\end{equation}

\subsection{The spectrum in perturbation theory}

\subsubsection{Small $|z|$}

In order to explain the structure of this spectrum we will use perturbation theory for small values of the noncommutativity parameters. For convenience, we take as unperturbed Hamiltonian $H_0$ and perturbation $V$ the operators given by
\begin{equation}\label{S21}
    \begin{array}{c} \displaystyle
      2MH_0=2 eB \left[ a^\dagger a + \frac{1}{2} \right]+2M\kappa s_0+\bar{z}z\left( {s_0}^2-{\mathbf{s}}^2\right)\,,
     \\ \\ \displaystyle
     2 M V= \sqrt{2eB}\left\{z a^\dagger s_+ + \bar{z} a s_- \right\}\,.
    \end{array}
\end{equation}

Since $H_0$ commutes with $L_0$, $s_0$ and $b^\dagger b$, the unperturbed eigenvectors and eigenvalues are given by
\begin{equation}\label{S22}
   \begin{array}{c}\displaystyle
      \Psi_{n,n_b ,m}=\left| n, n_b  \right\rangle \otimes \left| \lambda , m \right\rangle
      \,, \quad H_0   \Psi_{n,n_b ,m} = E_{n,m}^{(0)} \Psi_{n,n_b ,m}\,,
      \\ \\ \displaystyle
     E_{n,m}^{(0)} =\frac{1}{2 M} \left\{ 2eB\left(n+\frac{1}{2}\right)+2M\kappa m
     +\bar{z}z\left( {m}^2-\lambda\right) \right\}\,,
   \end{array}
\end{equation}
degenerated in the index $n_b $.

Since $V$ commutes with $b^\dagger b$, the first order corrections to the eigenvalues in perturbation theory are simply given by
\begin{equation}\label{S23}
    E_{n,m}^{(1)} = \left(\Psi_{n,n_b ,m} ,V \Psi_{n,n_b ,m}\right)=0\,,
\end{equation}
and are all vanishing.

The second order corrections are given by
\begin{equation}\label{S24}
    E_{n,m}^{(2)} = {\sum_{n',m'}}'\frac{\left| \left(\Psi_{n',m',n_b } ,V \Psi_{n,m,n_b }\right)\right|^2}{E_{n,m}^{(0)}-E_{n',m'}^{(0)}}\,,
\end{equation}
where the term with $n'=n$ y $m'=m$ is excluded from the series. From (\ref{S21}) we get
\begin{equation}\label{S25}
     \begin{array}{c} \displaystyle
       \left(\Psi_{n',n_b ,m'} ,2MV \Psi_{n,n_b ,m}\right)= \sqrt{2eB} z \sqrt{n+1} \sqrt{m(m+1)-\lambda}\, \delta_{n',n+1}\delta_{m',m+1} +
       \\ \\ \displaystyle
       + \sqrt{2eB} \bar{z} \sqrt{n} \sqrt{m(m-1)-\lambda}\, \delta_{n',n-1}\delta_{m',m-1}\,,
     \end{array}
\end{equation}
from which it follows that
\begin{equation}\label{S26}
   \begin{array}{c} \displaystyle
      E_{n,m}^{(2)} = \frac{-1}{2M}\, \frac{|z|^2(n+1)[m(m+1)-\lambda]}{1+\frac{M\kappa}{eB}+\frac{|z|^2}{2eB}(2m+1)}+
     \frac{1}{2M}\, \frac{|z|^2n[m(m-1)-\lambda]}{1+\frac{M\kappa}{eB}+\frac{|z|^2}{2eB}(2m-1)}=
     \\ \\ \displaystyle
     =- \frac{|z|^2}{2M}\left\{ 2nm+\left[m(m+1)-\lambda \right] \right\} + O\left(|z|^3\right)\,.
   \end{array}
\end{equation}

Then, up to second order in $|z|$, we get for the eigenvalues
\begin{equation}\label{S27}
   \begin{array}{c} \displaystyle
      E_{n,m}=\frac{eB}{M}\left(n+\frac{1}{2}\right)+\kappa m+\frac{|z|^2}{2 M} \left( {m}^2-\lambda\right)  - \frac{|z|^2}{2M}\left\{ 2nm+\left[m(m+1)-\lambda \right] \right\}+  O\left(|z|^3\right)=
      \\ \\ \displaystyle
       =\frac{eB-|z|^2 m}{M}  \left(n+\frac{1}{2}\right)+\kappa m+  O\left(|z|^3\right)\,,
   \end{array}
\end{equation}
for any unitary \emph{irrep} of $SL(2,\mathbb{R})$.  Notice that, at this order and for each $m$, these are the Landau levels for an \emph{effective magnetic field} linearly dependent on $m$, rigidly shifted by the $\kappa m$ term. Notice also that the dominant term in the $\theta$ parameter is quadratic.

For example, considering again the unitary \emph{irrep} of the discrete series with $k=\frac{1}{2}$ and $m\leq -k$ we get (up to $O\left(|z|^3\right)$ terms)
\begin{equation}\label{S28}
    \begin{array}{c}
      E_{0,-\frac{1}{2}}=\frac{eB+\frac{|z|^2}{2}}{M}  \left(\frac{1}{2}\right)-\frac{\kappa}{2}\,, \quad
       E_{0,-\frac{3}{2}}=\frac{eB+\frac{3|z|^2}{2}}{M}  \left(\frac{1}{2}\right)-\frac{3\kappa}{2}\,, \quad
       E_{0,-\frac{5}{2}}= \frac{eB+\frac{5|z|^2}{2}}{M}  \left(\frac{1}{2}\right)-\frac{5\kappa}{2}\,,
       \cdots
       \\ \\
     E_{1,-\frac{1}{2}}=\frac{eB+\frac{|z|^2}{2}}{M}  \left(1+\frac{1}{2}\right)-\frac{\kappa}{2}\,, \quad
       E_{1,-\frac{3}{2}}=\frac{eB+\frac{3|z|^2}{2}}{M}  \left(1+\frac{1}{2}\right)-\frac{3\kappa}{2}\,, \quad
       E_{1,-\frac{5}{2}}=\frac{eB+\frac{5|z|^2}{2}}{M}  \left(1+\frac{1}{2}\right)-\frac{5\kappa}{2}\,,
       \cdots
            \\ \\
     E_{2,-\frac{1}{2}}= \frac{eB+\frac{|z|^2}{2}}{M}  \left(2+\frac{1}{2}\right)-\frac{\kappa}{2} \,, \quad
       E_{2,-\frac{3}{2}}= \frac{eB+\frac{3|z|^2}{2}}{M}  \left(2+\frac{1}{2}\right)-\frac{3\kappa}{2}\,, \quad
       E_{2,-\frac{5}{2}}= \frac{eB+\frac{5|z|^2}{2}}{M}  \left(2+\frac{1}{2}\right)-\frac{5\kappa}{2}\,,
       \cdots
    \end{array}
\end{equation}
in complete agreement with Eqs.\ (\ref{S18}-\ref{S19}).

\subsubsection{Large $|z|$}

We will also consider the large NC parameters limit in perturbation theory. So, we now take as unperturbed Hamiltonian the operator
\begin{equation}\label{L1}
    \mathcal{H}_0:=\frac{\bar{z}z}{2M}\left( {s_0}^2 -  {\mathbf{s}}^2\right)
\end{equation}
and as perturbation
\begin{equation}\label{L2}
    \mathcal{V}:=\kappa s_0 + \frac{\sqrt{2eB}}{2M} \left( z a^\dagger s_+ \bar{z} a s_- \right) + \frac{eB}{M} \left( a^\dagger a + \frac{1}{2}\right)\,.
\end{equation}
The eigenvectors and eigenfunctions of $\mathcal{H}_0$ are given by
\begin{equation}\label{L3}
    \begin{array}{c}\displaystyle
    \chi_{n,n_b,m}:= \left|n, n_b \right\rangle \otimes \left|\lambda, m \right\rangle\,,
      \\\\ \displaystyle
      \mathcal{E}_{n,m}^{(0)}=  \frac{\bar{z}z}{2M}\left( {m}^2 -  \lambda\right)\,,
    \end{array}
\end{equation}
which depend only on $m$ and are degenerate in $n$ and $n_b$.

Since both $\mathcal{H}_0$ and $\mathcal{V}$ commute with $b^\dagger b$, we can refer to the subspace with definite $n_b$, and consider only the degeneracy in $n$. The first order correction to the eigenvalues in perturbation theory are given by the matrix elements
\begin{equation}\label{L4}
    \left( \chi_{n',n_b,m}, \mathcal{V}  \chi_{n,n_b,m} \right) = \delta_{n',n} \left\{ \kappa m + \frac{eB}{M} \left( n+\frac{1}{2} \right) \right\}\,,
\end{equation}
which are already diagonal in $n$.

The ($O(|z|)$) second term in the right hand side of Eq.\ (\ref{L2}) contributes at second order of perturbation theory with an $O\left(\frac{eB}{M}\right)$ correction. Then,
\begin{equation}\label{L5}
    \mathcal{E}_{n,m}=  \frac{\bar{z}z}{2M}\left( {m}^2 -  \lambda\right)+\kappa m + O\left(\frac{eB}{M}\right)\,.
\end{equation}
Then, one sees that the noncommutativity parameters appear as a typical energy scale for the separation of successive series of Landau levels. For $|z|/M \gg 1$, only the states with the minimum value of $m^2$ will manifest al low energies.

\section{Dirac Particles}\label{Dirac}

The Dirac equation in 2+1-dimensions is
\begin{equation}\label{dd1}
    \left( \imath \gamma^\mu \partial_\mu -M \right)\Psi=0\,,
\end{equation}
where we take
\begin{equation}\label{dd2}
    \gamma^0=\sigma_3\,, \quad \gamma^1=-\imath\sigma_2\,, \quad \gamma^2=\imath\sigma_1\,,
\end{equation}
which satisfy $\left[ \gamma^\mu , \gamma^\nu \right]=2 g^{\mu\nu}$ with $\left(g^{\mu\nu}\right)={\rm diag}\left(1,-1,-1\right)$. From (\ref{dd1}) we get the Hamiltoniano $H=\alpha_i\p_i+M \beta$, where  $\alpha_1=-\sigma_1$, $\alpha_2=-\sigma_2$ and $\beta=\sigma_3$.

In the presence of an external electromagnetic field, minimal coupling requires to change
$p_\mu\rightarrow p_\mu -e A_\mu$. So, the Hamiltonian becomes
\begin{equation}\label{dd3}
    H=\alpha_i\left(p_i-e A_i\right)- e A_0 + M \beta\,.
\end{equation}

We consider again a constant magnetic field perpendicular to the plane of the system. Then, we take again $A_0=0$ and $A_i(\mathbf{x})=-\frac{B}{2} \epsilon_{i j} x^j$, obtaining
\begin{equation}\label{dd4}
    \begin{array}{c} \displaystyle
      H
     = \alpha_1  \sqrt{e B }\, q +\alpha_2 \sqrt{e B}\, p  + M \beta =
     \\ \\ \displaystyle
    = \left(
                                  \begin{array}{cc}
                                    M & -\sqrt{2e B }\, a^\dagger \\
                                    -\sqrt{2e B }\, a & -M \\
                                  \end{array}
                                \right)   \,,
    \end{array}
\end{equation}
in terms of the operators defined in Eq.\ (\ref{S2}) and (\ref{S5}).

Taking into account that
\begin{equation}\label{dd5}
   \begin{array}{c}\displaystyle
      \left[H , L_0 \right] =   \left[H , b^\dagger b-a^\dagger a \right]
     = \sqrt{2e B } \left(
     \begin{array}{cc}
     0 &  -a^\dagger \\
     a  & 0 \\
     \end{array}
     \right)
   \end{array}
\end{equation}
and
\begin{equation}\label{dd6}
     \left[H , \sigma_3 \right] =-\sqrt{2e B } \left[a^\dagger \sigma_+ + a \sigma_- , \sigma_3 \right]
     = 2 \sqrt{2e B } \left(
                                                                                          \begin{array}{cc}
                                                                                            0 &  a^\dagger \\
                                                                                          -a  & 0 \\
                                                                                          \end{array}
                                                                                        \right)\,,
\end{equation}
we conclude that $H$ commutes with $J_0:=L_0+\frac{\sigma_3}{2}$. Consequently, we can write the eigenvectors of $H$ and $J_0$ as
\begin{equation}\label{dd7}
    \psi_{n,n_b }  = \left(
    \begin{array}{c}
    C_1 \left|n,n_b  \right\rangle \\
    C_2 \left|n-1,n_b  \right\rangle \\
    \end{array}
    \right)\,,
\end{equation}
with $n\geq 1$. Indeed, we have
\begin{equation}\label{dd8}
    J_0 \psi_{n,n_b }  =\left(
    \begin{array}{c}
    C_1 \left(L_0+\frac{1}{2} \right) \left|n,n_b  \right\rangle \\
    C_2 \left(L_0-\frac{1}{2} \right) \left|n-1,n_b  \right\rangle \\
    \end{array}
    \right)= j_0 \psi_{n,n_b } \,,
\end{equation}
with eigenvalue $j_0=n_b -n+\frac{1}{2}$.

On the other hand, $(H-E) \psi_{n,n_b }  =0$ implies that
\begin{equation}\label{dd10}
     \left(
       \begin{array}{cc}
         M-E & -\sqrt{2e B n}  \\
         -\sqrt{2e B n}  & -(M+E) \\
       \end{array}
     \right) \left(
               \begin{array}{c}
                 C_1 \\
                 C_2 \\
               \end{array}
             \right) =  \left(
               \begin{array}{c}
                 0 \\
                0 \\
               \end{array}
             \right)\,.
\end{equation}
Nontrivial solutions require that
\begin{equation}\label{dd11}
    E^2-M^2-2eBn=0 \quad \Rightarrow \quad E_{\pm,n}=\pm\sqrt{M^2+2eBn}\,,
\end{equation}
and
\begin{equation}\label{ddd11}
    C_2=\frac{E_{\pm,n}-M}{\sqrt{2e B n} } \, C_1\,,
\end{equation}
both independent of $n_b $. Then, the eigenvectors are
\begin{equation}\label{sol-mas-menos}
    \psi_{\pm,n,n_b }= \left(
    \begin{array}{c}
    {\sqrt{2e B n } } \left|n,n_b  \right\rangle \\
    \left[ -M \pm\sqrt{M^2+2eBn} \right] \left|n-1,n_b  \right\rangle \\
    \end{array}
    \right)\,,
\end{equation}
with $n\geq 1$, degenerate in the index $n_b $.

There is another solution for $n=0$, given by
\begin{equation}\label{dd12}
   \psi_{0,n_b } = \left(
   \begin{array}{c}
   \left|0,n_b  \right\rangle \\
   0 \\
   \end{array}
   \right)\,,
\end{equation}
with $j_0=n_b +1/2$ and $E_0=M$, also degenerate in $n_b $.

\subsection{Extension to the noncommutative space}

We adopt as Hamiltonian of this system the Hermitian operator
\begin{equation}\label{dd16}
   \begin{array}{c} \displaystyle
      \hat{H}=\alpha_i\left(\hat{p}_i-e A_i(\hat{\mathbf{x}})\right)+\kappa s_0+ M \beta=
      \\ \\ \displaystyle
     =H \otimes \mathbf{1}+\kappa \left(\alpha_i \otimes s_i + \mathbf{1}_2\otimes s_0\right) - \theta \frac{e B}{2}  \epsilon_{i j}\, \alpha_i \otimes s_j\,,
   \end{array}
\end{equation}
where the term $\left(\kappa  \mathbf{1}_2\otimes s_0\right)$ comes from the shift of $p_0$ in Eq.\ (\ref{shift}). This Hamiltonian can also be written as
\begin{equation}\label{dd17}
    \begin{array}{c} \displaystyle
      \hat{H}= H \otimes \mathbf{1} + \kappa \mathbf{1} \otimes s_0-z \sigma_- \otimes s_+
                                        -\bar{z} \sigma_+ \otimes s_-\,,
    \end{array}
\end{equation}
where $\sigma_\pm=\frac{\sigma_1\pm \imath \sigma_2}{2}$, $s_\pm:=s_1\pm \imath s_2$ (See Eq.\ (\ref{Xmasmenos})).
% and $\kappa=\frac{z+\bar{z}}{2}$.

This Hamiltonian has a symmetry generated by
\begin{equation}\label{dd18}
    J:=\left( L_0+\frac{1}{2}\, \sigma_3 \right) \otimes \mathbf{1} + \mathbf{1}_2 \otimes s_0\,.
\end{equation}
Indeed (See Eq.\ (\ref{17})),
\begin{equation}\label{dd19}
    \begin{array}{c} \displaystyle
       \left[J , \hat{H} \right]
        = - \frac{z}{2}  \left[ \sigma_3 , \sigma_- \right] \otimes s_+
         - \frac{\bar{z}}{2}  \left[ \sigma_3 , \sigma_+ \right] \otimes s_- -
           \\ \\ \displaystyle
         -z  \sigma_- \otimes \left[ s_0, s_+\right]
                                        -\bar{z}  \sigma_+ \otimes  \left[ s_0,s_-\right]=0\,.
    \end{array}
\end{equation}

Moreover, $\left[ b^\dagger b, \hat{H} \right]=0$. Then, the energy eigenvalues are also degenerate in $n_b $.

\medskip

Given a unitary \emph{irrep} of $SL(2,\mathbb{R})$, we can choose a complete system of orthogonal vectors in the subspace of the Hilbert space characterized by given values of $n_b $ y $j$ (the eigenvalue of $J$) as
\begin{equation}\label{dd20}
    \left\{
    \begin{array}{c} \displaystyle
      \psi_{n,\uparrow}= \left(
                           \begin{array}{c}
                           \left|n,n_b \right\rangle \\
                             0 \\
                           \end{array}
                         \right) \otimes \left| \lambda, j-n_b -\frac{1}{2}+n \right\rangle \,,
      \\ \\ \displaystyle
      \psi_{n,\downarrow}= \left(
                           \begin{array}{c}
                            0 \\
                           \left|n,n_b \right\rangle \\
                           \end{array}
                         \right) \otimes \left| \lambda, j-n_b +\frac{1}{2}+n \right\rangle \,.
\end{array}
    \right.
\end{equation}
Indeed, in both cases, the eigenvalue of $J$ is $\left(n_b -n\pm\frac{1}{2}\right)+\left(j-n_b \mp\frac{1}{2}+n\right)=j$.

Let us point out that
\begin{equation}\label{dd22}
    \begin{array}{c} \displaystyle
      \left[  (H-E) \otimes \mathbf{1} + \kappa \mathbf{1}_2 \otimes s_0 \right] \psi_{n,\uparrow}  = \left[ M-E+\kappa\left(j-n_b -\frac{1}{2}+n\right) \right]  \psi_{n,\uparrow} -
      \sqrt{2eB n} \, \psi_{n-1,\downarrow}  \,,
     \end{array}
\end{equation}
and
\begin{equation}\label{dd23}
    \begin{array}{c} \displaystyle
      \left[  (H-E) \otimes \mathbf{1} + \kappa \mathbf{1}_2 \otimes s_0 \right] \psi_{n,\downarrow}  = \left[ -M-E+\kappa\left(j-n_b +\frac{1}{2}+n\right) \right]  \psi_{n,\downarrow} -
      \sqrt{2eB (n+1)} \, \psi_{n+1,\uparrow}  \,.
     \end{array}
\end{equation}
Moreover,
\begin{equation}\label{dd24}
    \begin{array}{c} \displaystyle
      \sigma_- \otimes s_+ \psi_{n,\uparrow}=
      \sqrt{\left( j-n_b +n \right)^2-\left(\lambda +\frac{1}{4} \right)}\, \psi_{n,\downarrow}\,,
    \end{array}
\end{equation}
and
\begin{equation}\label{dd25}
    \begin{array}{c} \displaystyle
       \sigma_+ \otimes s_- \psi_{n,\downarrow}=
       \sqrt{\left( j-n_b +n \right)^2-\left(\lambda +\frac{1}{4} \right)}\, \psi_{n,\uparrow} \,,
    \end{array}
\end{equation}
while
\begin{equation}\label{dd26}
     \sigma_+ \otimes s_- \psi_{n,\uparrow}=0=\sigma_- \otimes s_+ \psi_{n,\downarrow}\,.
\end{equation}

\medskip

If we propose the following development for the Hamiltonian eigenvectors,
\begin{equation}\label{dd21}
    \begin{array}{c} \displaystyle
      \Psi= \sum_{n=0}^{\infty} \left( C_{n}  \psi_{n,\uparrow} + D_n  \psi_{n,\downarrow} \right)=
      \\ \\ \displaystyle
      = \sum_{n=0}^{\infty}   \left(
                           \begin{array}{c}
                           C_{n} \left|n,n_b \right\rangle  \otimes \left| \lambda, j-n_b -\frac{1}{2}+n \right\rangle \\ \\
                             D_n  \left|n,n_b \right\rangle  \otimes \left| \lambda, j-n_b +\frac{1}{2}+n \right\rangle  \\
                           \end{array}
                         \right)\,,
    \end{array}
\end{equation}
the condition $\left( \hat{H}-E \right) \Psi=0$ straightforwardly leads to the recurrence relations
\begin{equation}\label{dd28}
    \left\{
     \begin{array}{c} \displaystyle
       C_n \left[ M-E+\kappa\left(j-n_b -\frac{1}{2}+n\right) \right] - D_{n-1}  \sqrt{2eB n}
      -D_n \bar{z}  \sqrt{\left( j-n_b +n \right)^2-\left(\lambda +\frac{1}{4} \right)}=0\,,
      \\ \\ \displaystyle
      D_n \left[ -M-E+\kappa\left(j-n_b +\frac{1}{2}+n\right) \right] - C_{n+1}  \sqrt{2eB (n+1)}
      -C_n z  \sqrt{\left( j-n_b +n \right)^2-\left(\lambda +\frac{1}{4} \right)}=0\,,
     \end{array}
     \right.
\end{equation}
for  $n\geq 0$ and where $D_{-1}:=0$. Notice that the solution depends on $j$ and $n_b $ only through the difference $j-n_b $.

It is easy to verify that the limit $\kappa,\theta\rightarrow0$ reproduces the results in Eq.\ (\ref{dd11}), (\ref{sol-mas-menos}) and (\ref{dd12}).

\medskip

The problem of getting the Hamiltonian eigenvectors appears to be more difficult than in the case of Schr\"{o}dinger particles. But, as before, for certain unitary \emph{irrep}'s of $SL(2,\mathbb{R})$ it reduces to a matricial eigenvalue problem.

Indeed, if we consider again an \emph{irrep} of $SL(2,\mathbb{R})$ in the discrete series, characterized by $\lambda=k(k-1)$ and $m\leq -k$, one can see that $m=j-n_b +n-\frac{1}{2}\leq -k \quad \Rightarrow \quad 0\leq n \leq n_b -j-k+\frac{1}{2}$.

For example, taking $k=\frac{1}{2}$ with $j-n_b =0$ we simply get
\begin{equation}\label{dd30}
    C_0 \left(-E-\frac{\kappa }{2}+M\right)=0\,.
\end{equation}
Then, $E=M-\frac{\kappa }{2}$ and $\Psi \sim \psi_{0,\uparrow}$.

For $j-n_b =-1$, the eigenvalues are the zeroes of the determinant
\begin{equation}\label{dd31}
    \left|
\begin{array}{ccc}
 -\frac{3 \kappa }{2}+M-E & 0 & -\bar{z} \\ \\
 0 & -\frac{\kappa }{2}+M-E & - \sqrt{2B e} \\ \\
 -z & - \sqrt{2B e} & -\frac{\kappa }{2}-M-E \\
\end{array}
\right|=0\,,
\end{equation}
which have a rather involved expression as roots of a polynomial of degree 3.

\subsection{The spectrum in perturbation theory}

\subsubsection{Small $|z|$}

In order to explain the structure of the spectrum for small noncommutative parameters we will use again perturbation theory. We take $H_0:=H \otimes \mathbf{1} + \kappa \mathbf{1} \otimes s_0$ as the unperturbed Hamiltonian and $V:=-z \sigma_- \otimes s_+ -\bar{z} \sigma_+ \otimes s_-$ as the perturbation.

Since $H_0$ commutes with $\left(L_0+\frac{1}{2}\, \sigma_3\right)$ and with $s_0$, we can take as unperturbed normalized eigenvectors and eigenvalues
\begin{equation}\label{dd32}
    \begin{array}{c} \displaystyle
    \Psi_{0,n_b ,m}=\left(
                 \begin{array}{c}
                   \left| 0, n_b  \right\rangle \\ \\
                   0 \\
                 \end{array}
               \right) \otimes \left| \lambda, m\right\rangle\,, \quad E_{0,m}^{(0)}=M+ \kappa m\,,
      \\ \\ \displaystyle
      \Psi_{\pm,n,n_b ,m}=C_{\pm,n}\left(
                 \begin{array}{c}
                  \sqrt{2eBn} \left| n, n_b  \right\rangle \\ \\
                   \left[ M\mp \sqrt{M^2+2eBn} \right]\left| n-1, n_b  \right\rangle \\
                 \end{array}
               \right) \otimes \left| \lambda, m\right\rangle\,, \quad E_{\pm,n,m}^{(0)}=\pm\sqrt{M^2+2eBn}+ \kappa m\,,
    \end{array}
\end{equation}
degenerate in $n_b \in \mathbb{N}\cup \left\{0\right\}$, with
\begin{equation}\label{dd33}
    C_{\pm,n}=\left\{2\left( M^2+2eBn \right)\mp 2 M\sqrt{M^2+2eBn } \right\}^{-\frac{1}{2}}\,.
\end{equation}

Since $\left[b^\dagger b, V\right]=0$, we can refer to the subspace with a given $n_b$. Then, the first order corrections to the energies in perturbation theory are all vanishing. Indeed, they are simply given by
\begin{equation}\label{dd34}
    \left(\Psi_{0,n_b ,m} , V \Psi_{0,n_b ,m}\right)=0=\left(  \Psi_{\pm,n,n_b ,m} , V   \Psi_{\pm,n,n_b ,m}\right)\,.
\end{equation}

On the other hand,
\begin{equation}\label{dd35}
   \begin{array}{c} \displaystyle
      \left( \Psi_{s',n',n_b ,m'} , V \Psi_{0,n_b ,m}\right)=-\delta_{n',1}\delta_{m',m+1} z C_{s',n'} \left[ M-s' \sqrt{M^2+2eB} \right]\sqrt{m(m+1)-\lambda}\,,
      \\ \\ \displaystyle
      E_{0,m}^{(0)} -  E_{s',n',m'}^{(0)}=M-s' \sqrt{M^2+2eBn'}+\kappa(m-m')\,,
   \end{array}
\end{equation}
and
\begin{equation}\label{dd36}
   \begin{array}{c} \displaystyle
      \left( \Psi_{s',n',n_b ,m'} , V \Psi_{s,n,n_b ,m}\right) =
       \\ \\ \displaystyle
     =  -C_{s',n'}C_{s,n} \left\{\delta_{n',n+1}\delta_{m',m+1} z \left[ M-s' \sqrt{M^2+2eB(n+1)}\right]\sqrt{2eBn} \sqrt{m(m+1)-\lambda} +\right.
      \\ \\ \displaystyle
      \left. + \delta_{n',n-1}\delta_{m',m-1} \bar{z} \sqrt{2eB(n-1)}  \left[ M-s \sqrt{M^2+2eBn}\right]\sqrt{m(m-1)-\lambda}  \right\}\,,
            \\ \\ \displaystyle
      E_{s,n,m}^{(0)} -  E_{s',n',m'}^{(0)}=s \sqrt{M^2+2eBn}-s' \sqrt{M^2+2eBn'}+\kappa(m-m')\,,
   \end{array}
\end{equation}
From Eqs.\ (\ref{dd35}) and (\ref{dd36}) one can easily compute the $\left(O(|z|^2)\right)$ second order corrections to the energies.

Therefore, for any \emph{irrep} of $SL(2,\mathbb{R})$ and to first order in $|z|$, the energy eigenvalues are given in Eq.\ (\ref{dd32}). Notice that, as in the  the case of Schr\"{o}dinger particles, they show a shift linear in $m$ and they do not depend on $\theta$ at first order. This is also in agreement with Eq.\ (\ref{dd30}) and (\ref{dd31}), up to $O\left( |z|^2 \right)$ terms.

\subsubsection{Large $|z|$}

In the large $|z|$ limit, we take as unperturbed Hamiltonian the operator
\begin{equation}\label{LL1}
    \mathcal{H}_0: = \kappa \mathbf{1} \otimes s_0-z \sigma_- \otimes s_+
    -\bar{z} \sigma_+ \otimes s_-
\end{equation}
and as the perturbation
\begin{equation}\label{LL2}
    \mathcal{V}:=  H \otimes \mathbf{1}=
    M \sigma_3\otimes \mathbf{1} -\sqrt{2e B }
    \left[a^\dagger \sigma_+ + a \sigma_- \right] \otimes \mathbf{1} \,,
\end{equation}

Since $\left[\mathcal{H}_0 , \frac{1}{2}\, \sigma_3+s_0 \right]=0$, one can see that the normalized eigenvectors of $\mathcal{H}_0$ are given by
\begin{equation}\label{LL3}
    \Phi_{n,n_b,j,\pm}= \left| n,n_b\right\rangle \otimes \left(
    \begin{array}{c}
    c_1(j,\pm) \left| \lambda, j-\frac{1}{2} \right\rangle \\ \\
     c_2(j,\pm) \left| \lambda, j+\frac{1}{2} \right\rangle\\
     \end{array}
     \right)\,,
\end{equation}
where $j$ is the eigenvalue of $\left(\frac{1}{2}\, \sigma_3+s_0\right)$ and
\begin{equation}\label{LL4}
    \begin{array}{c} \displaystyle
      c_1(j,\pm)=-\frac{\sqrt{2}\,\bar{z}}{\kappa} \frac{\sqrt{\left[j^2-\left(\lambda +\frac{1}{4}\right)\right]}}
      {\sqrt{1+4 \gamma  \left[j^2-\left(\lambda +\frac{1}{4}\right)\right]\pm
   \sqrt{1 + 4 \gamma  \left[j^2-\left(\lambda +\frac{1}{4}\right)\right] }}}\,,
       \\ \\ \displaystyle
      c_2(j,\pm)=\frac{2 {\mathcal{E}^{(0)}_{j,\pm}}+\kappa(1 -2 j)}
      {\sqrt{2} \kappa  \sqrt{1+4 \gamma  \left[j^2-\left(\lambda
   +\frac{1}{4}\right)\right]\pm \sqrt{1+4 \gamma
   \left[j^2-\left(\lambda +\frac{1}{4}\right)\right]}}}\,,
    \end{array}
\end{equation}
with $\gamma:=\bar{z} z/\kappa^2$ and ${\mathcal{E}^{(0)}_{j,\pm}}$ the corresponding eigenvalues,
\begin{equation}\label{LL5}
    {\mathcal{E}^{(0)}_{j,\pm}}:=\kappa  \left(j\pm \frac{1}{2}
    \sqrt{1+4 \gamma \left[j^2-\left(\lambda +\frac{1}{4}\right) \right]}\right)\,,
\end{equation}
degenerate in  the indices $n$ and $n_b$.

The corrections to the energies at first order in perturbation theory get contributions from the first term in the right hand side of Eq.\ (\ref{LL2}) only and are determined by the matrix elements
\begin{equation}\label{LL6}
    \begin{array}{c} \displaystyle
      {\mathcal{E}^{(1)}_{j,\pm}}=
      \left(\Phi_{n',n_b',j,\pm} , \mathcal{V}  \Phi_{n,n_b,j,\pm} \right)
    = M \delta_{n',n}\delta_{n_b',n_b} \left\{\left|c_1(j,\pm)\right|^2 -
    \left|c_2(j,\pm)\right|^2\right\} =
    \\ \\ \displaystyle
      = \frac{\mp M}{\sqrt{1+4 \gamma
      \left[j^2-\left(\lambda +\frac{1}{4}\right)\right]}} \, \delta_{n',n}\delta_{n_b',n_b}\,,
    \end{array}
\end{equation}
which are already diagonal.

It can be easily seen that the the second order corrections in perturbation theory get an $O(M^2/|z|)$ contribution from the first term in the right hand side of Eq.\ (\ref{LL2}) and $O(e B/|z|)$ contributions from the second term in the right hand side of that equation.

Then, also in this model the noncommutativity parameters appear as a typical energy scale for the separation of successive series of Landau levels and, at low energies, only the states with the minimum value of $j$ would manifest.

\section{Conclusions}\label{conclusion}

%%%%%%%%%%%%%%%%%%%%%%%%%%%%%%%%%%%%%%%%%%%%%%%%%%%%%%

In this article we have considered models of Schr\"{o}dinger and Dirac particles living in a space-time with a nonstandard noncommutativity, both in coordinates and momenta. This noncommutativity was induced by deforming the canonical commutators by terms proportional to the generators in a unitary irreducible representation of the Lorentz group in the 2+1-dimensional Minkowski space, isomorphic to $SL(2,\mathbb{R})/\mathbb{Z}_2$. Since this is a noncompact Lie group, its unitary \emph{irrep}'s are not of finite dimension.

Taking into account that $SL(2,\mathbb{R})$ is a three-dimensional Lie group, we have realized this deformation of the Heisenberg algebra by shifting canonical coordinates and momenta with terms proportional to the generators in the unitary \emph{irrep} considered, a kind of non-Abelian Bopp's shift. In particular, the shift in momenta can also be interpreted as the introduction of a non-Abelian magnetic field.

Consequently, the number of dynamical variables is enlarged and the Hilbert space gets the structure of a direct product, one factor for the state vectors of the ordinary system in the normal space and the other for the component of the state vectors in the representation space of this \emph{irrep} of $SL(2,\mathbb{R})$.

We have shown that total generators of the Lorentz transformations can be constructed which correctly transform all the operators, thus realizing the Lie algebra $sl(2,\mathbb{R})$ on the Hilbert space of the quantum-mechanical system.

In this framework, we have considered modified Hamiltonians obtained through this non-Abelian Bopp's shift of the dynamical variables from the Hamiltonians of the Landau problem for both Schr\"{o}dinger and Dirac particles. We have analyzed these models for both discrete and continuous classes of \emph{irrep}'s of $sl(2,\mathbb{R})$. In general, the eigenvalue problem leads to an infinite recursion relation for the coefficients in the development of the eigenvectors in terms of a conveniently chosen bases of the Hilbert space, although for certain \emph{irrep}'s it reduces to a matricial problem. The spectrum of these models have been studied also in perturbation theory, both for small and large noncommutativity parameters $z= \kappa+\imath \theta \frac{eB}{2}$.

In the case of a Schr\"{o}dinger particle, Eq.\ (\ref{S27}) shows that for small $|z|$ and for any \emph{irrep} of $SL(2,\mathbb{R})$ there is a series of Landau levels, one for each eigenvalue $m$ of $s_0$, rigidly shifted by a term proportional to $m$ and with a second order correction to the effective magnetic field. On the other hand, in the limit of large $|z|$ Eqs.\ (\ref{L5})  and (\ref{L4}) show that the noncommutativity parameters appear as a typical energy scale for the separation of successive series of Landau levels and that, at low energies, only the Landau levels with the minimum value of $m^2$ manifest. Similar conclusions have been obtained for the model of a Dirac particle.

Let us mention that, contrary to the case of conventional NC Quantum Mechanics, we have found no constraint between the parameters referring to no-commutativity in coordinates and momenta. Rather, with a nonvanishing magnetic field $B$, both $\kappa$ and $\theta$ play a similar role (although there are no linear in $\theta$ contributions to the eigenvalues).

Notice that the structure of the Hilbert space as a direct product leads, in the $|z|\to 0$ limit, to an infinite degeneracy additional to the usual degeneracy of the Landau problem. In this sense, the noncommutative models here considered do not reduce to the original ones in this limit. Indeed, the modified Hamiltonian $\hat{H}$ takes the form $H \otimes \mathbf{1}_{irrep}$ in this limit, being then diagonal in the factor space of the representation of the group. Therefore, these models do not correspond to just a smooth deformation of the commutative ones. On the other hand, as previously mentioned, in the $|z| \to \infty$ limit only the lowest excitations in this additional factor of the Hilbert space would be detected in the low energy limit, with no evidence of the higher levels.

It is worthwhile to remark that the representation space of the unitary \emph{irrep}'s of $SL(2,\mathbb{R})$ can be explicitly realized in terms of spaces of square-integrable functions: functions defined on the unit circle for the continuous classes of  \emph{irrep}'s and analytic functions on the unit open disk for the discrete classes of  \emph{irrep}'s, as discussed in Appendix \ref{irrep-unit}. Therefore, the examples studied in this article can also be considered as equivalent to models of quantum mechanical particles living in a space with an additional compact dimension, with the parameter $\kappa$ playing the role of the inverse of a typical length. Indeed, the non-Abelian Bopp's shift in Eq.\ (\ref{shift}) leads to a description of these systems in terms of the usual phase-space variables of a (commutative) (2+1)-dimensional space plus the generators of an \emph{irrep} of $SL(2,\mathbb{R})$, which are the dynamical variables adequate to describe its behavior in this additional dimension.

%%%%%%%%%%%%%%%%%%%%%%%%%%%%%%%%%%%%%%%%%%%%%%%%%%%%%%

\vskip 1cm

\noindent
\textbf{Acknowledgements}: J.G.\ acknowledge the Department of Physics of Pont.\ Univ.\ Cat\'{o}lica de Chile for the hospitality during the academic year 2011-2012. J.G.\ also thanks A.P.\ Polychronakos for his kind hospitality at the City College. F.V.\ acknowledge support from CONICET, Argentina. This work was partially supported by grants from CONICET (PIP 01787), ANPCyT (PICT-2011-0605) and UNLP (Proy.~11/X615), Argentina, and from FONDECYT (Grants 1095106, 1095217 and 1100777) and Proyecto Anillos ACT119, Chile.

%%%%%%%%%%%%%%%%%%%%%%%%%%%%%%%%%%%%%%%%%%%%%%%%%%%%%%%
\appendix

\section{The Lorentz group in 1+2-dimensions}\label{SL2R}

The Lorentz group in 1+2-dimensional Minkowski space \cite{Bargmann,Binegar,Jackiw}, $\mathbb{M}_3$, can be defined as the set of real linear transformations of $\mathbf{x}=(x^0,x^1,x^2)$, $\mathbf{x}'=L \mathbf{x}$, which leaves invariant the interval
\begin{equation}\label{2}
    s^2=\mathbf{x}^t \eta \mathbf{x}=\mathbf{x}^t L^t \eta L \mathbf{x}
\end{equation}
for all $\mathbf{x}$, where the metric $\left(\eta_{\mu \nu}\right) = {\rm diag}\left(+1,-1,-1\right)$. This means that
\begin{equation}\label{4}
    L^t \eta L = \eta  \quad \Rightarrow \quad \left({\rm det} L\right)^2=1\quad {\rm and}\quad
    \eta_{\alpha\beta} L^\alpha_{\ \mu} L^\beta_{\ \nu} =\eta_{\mu\nu}\,.
\end{equation}
Then,
\begin{equation}\label{5}
    {\rm det}\, L=\pm 1 \quad {\rm and} \quad L^0_{\ 0}\geq 1 \quad {\rm or}\quad L^0_{\ 0}\leq -1 \,.
\end{equation}

The connected part of the Lorentz group (the one containing the identity $\mathbf{1}_3$), $\mathcal{L}_+^\uparrow$, corresponds to the subgroup of transformations with ${\rm det}\, L=1$ and $L^0_{\ 0}\geq 1$. The other cosets of the group are obtained from $\mathcal{L}_+^\uparrow$ through the multiplication by the parity ($P:={\rm diag}\left(+1,-1,+1\right)$) and/or time-reversal ($T:={\rm diag}\left(-1,+1,+1\right)$) transformations.

\medskip

It is easy to see that $\mathcal{L}_+^\uparrow\approx SL(2,\mathbb{R})/\mathbb{Z}_2$. Indeed, one can establish a one-to-one correspondence between $\mathbb{M}_3$ and the space of real symmetric $2\times 2$ matrices through the relation
\begin{equation}\label{7}
    \sigma(\mathbf{x}):= x^0 \mathbf{1}_2+x^1 \sigma_3 + x^2 \sigma_1 =
    \left(
    \begin{array}{cc}
    x^0+x^1 & x^2 \\
    x^2 & x^0-x^1 \\
    \end{array}
    \right)=\sigma(\mathbf{x})^t
    % =\sigma(\mathbf{x})^\dagger
    \,,
\end{equation}
where $\sigma_1$ and $\sigma_3$ are the two real Pauli matrices.

Within this representation of Minkowski space, the interval is expressed as ${\rm det}\, \sigma(\mathbf{x})=\left(x^0\right)^2-\left(x^1\right)^2-\left(x^2\right)^2 =s^2$. Then, the Lorentz transformations are realized as $\sigma(\mathbf{x}) \rightarrow \Lambda \sigma(\mathbf{x}) \Lambda^t$ with real matrices $\Lambda$ such that ${\rm det}\, \Lambda =\pm 1$. These conditions define a Lie group whose connected part containing the identity $\mathbf{1}_2$ is $SL(2,\mathbb{R})$ (isomorphic to $SU(1,1)$).

Moreover, since the elements in the center of the group, $\left\{\mathbf{1}_2,-\mathbf{1}_2\right\}\approx\mathbb{Z}_2$, correspond to the same Lorentz transformation, we conclude that there exists a homomorphism $\phi:SL(2,\mathbb{R})\rightarrow \mathcal{L}_+^\uparrow$ which apply $\left\{+U,-U\right\}\rightarrow L$.
% with $U\in SL(2,\mathbb{R})$ and $L\in \mathcal{L}_+^\uparrow$.

The elements in $SL(2,\mathbb{R})$ can be written as
\begin{equation}\label{10}
    \Lambda=\left(
              \begin{array}{cc}
                a+c & b+d \\
               - b+d & a-c \\
              \end{array}
            \right)\,, \quad {\rm with}\quad a^2+b^2=1+c^2+d^2\geq 1\,.
\end{equation}
These elements can be parametrized as
\begin{equation}\label{12}
    \begin{array}{c}\displaystyle
      c=\sinh \alpha \, \cos \beta\,, \quad d=\sinh \alpha \, \sin \beta\,,
      \\ \\ \displaystyle
       a=\cosh \alpha \, \cos \gamma\,, \quad b=\cosh \alpha \, \sin \gamma\,,
    \end{array}
\end{equation}
with $\alpha\in\mathbb{R}$ and $\beta,\gamma\in[0,2\pi)$. Therefore, $SL(2,\mathbb{R})$ is a noncompact multiply connected 3-dimensional Lie group. As a consequence, the unitary irreducible representations of $SL(2,\mathbb{R})$ are not of finite dimension.

\medskip

Writing the elements of $SL(2,\mathbb{R})$ as $\Lambda=e^{ \imath A}$, one can see that a basis of the Lie algebra $sl(2,\mathbb{R})$ can be chosen as the set of matrices $\left\{X_0:=-\frac{1}{2}\,\sigma_2,X_1:=\frac{\imath}{2}\,\sigma_1,X_2:=\frac{\imath}{2}\,\sigma_3\right\}$, which satisfy the commutation relations
\begin{equation}\label{111}
    \left[X_\mu,X_\nu\right]=- \imath \epsilon_{\mu\nu\lambda} X^\lambda\,,
\end{equation}
where $ X^\mu=\eta^{\mu\nu} X_\nu$ and $\epsilon_{\mu\nu\lambda}$ totally antisymmetric with $\epsilon_{012}=1$. $X_0$ generates the rotations on the plane while $X_{1,2}$ correspond to the boosts in the spatial axis. The quadratic Casimir invariant is given by
\begin{equation}\label{122}
    \mathbf{X}^2:=\eta^{\mu\nu}X_\mu X_\nu={X_0}^2-{X_1}^2-{X_2}^2\,,
\end{equation}
which commutes with the generators $X_\mu$.

\subsection{Finite dimensional \emph{irrep}'s of $sl(2,\mathbb{R})$}\label{rep-matriciales}

Since $SL(2,\mathbb{R})$ is noncompact, its finite dimensional \emph{irrep}'s are not unitary. They can be constructed from the unitary \emph{irrep}'s  of $SU(2)$ in the following way.  The generator of rotations, $X_0$, is Hermitian in any \emph{irrep} and can be chosen as $X_0\rightarrow J_3$. Then, the other two generators are anti-Hermitian and can be taken as $X_1\rightarrow - \imath J_2$ y $X_2 \rightarrow \imath J_1$, where the $J_i\,, i=1,2,3$ are the generators of the unitary $j$-\emph{irrep}  of $su(2)$.

The $(2j+1)$-dimensional space representation is generated by the basis of vectors $\left\{ \left|j,m\right\rangle\,, m=-j,-j+1,\cdots,j-1,j \right\}$, and the Casimir operator reduces to $\mathbf{X}^2={J_3}^2-(-\imath J_2)^2-(\imath J_1)^2= \mathbf{J}^2=j(j+1) \mathbf{1}$, where $j(j+1) \geq 0$.

But, as previously discussed, we need the unitary representations of $SL(2,\mathbb{R})$, which are considered in the next Section.

\subsection{Unitary \emph{irrep}'s of $sl(2,\mathbb{R})$}\label{irrep-unit}

The unitary \emph{irrep}'s of $sl(2,\mathbb{R})$ are infinite-dimensional \cite{Bargmann}. They are generated by Hermitian operators  ${X_\mu}={X_\mu}^\dagger$ satisfying the commutation relations in Eq.\ (\ref{111}). In this case, the Casimir invariant can also take negative values.

Since $\left[X_\mu , \mathbf{X}^2 \right]=0$, let us consider a simultaneous normalized eigenvector of ${\mathbf{X}^2}$  and  $X_0$,
\begin{equation}\label{16}
    {\mathbf{X}^2} \left|\lambda,m\right\rangle = \lambda \left|\lambda,m\right\rangle\,, \quad X_0 \left|\lambda,m\right\rangle = m \left|\lambda,m\right\rangle\,,
\end{equation}
where $\lambda\in \mathbb{R}$ and $m$ takes integer or half-integer values.

If we define
\begin{equation}\label{Xmasmenos}
    X_\pm:= X_1 \pm \imath X_2\,, \quad {\rm with}\quad  {X_\pm}^\dagger = {X_\mp}\,,
\end{equation}
we get
\begin{equation}\label{17}
     \left[X_0,X_\pm\right]=\pm  X_\pm\,,\quad  \left[X_+,X_-\right]=-2  X_0\,.
\end{equation}
Then,
\begin{equation}\label{18}
   \begin{array}{c}\displaystyle
      X_0\left(X_\pm \left|\lambda,m\right\rangle \right)= X_\pm \left(X_0 \pm 1\right) \left|\lambda,m\right\rangle= (m\pm1)\left(X_\pm \left|\lambda,m\right\rangle \right)\,,
      \\ \\ \displaystyle
     \mathbf{X}^2 \left(X_\pm \left|\lambda,m\right\rangle \right)= \lambda \left(X_\pm \left|\lambda,m\right\rangle \right)\,.
   \end{array}
\end{equation}

Taking into account that
\begin{equation}\label{19}
    X_\pm X_\mp=X_0\left(X_0\mp 1\right) - \mathbf{X}^2\,,
\end{equation}
we conclude that
\begin{equation}\label{20}
    \left\| X_\mp \left|\lambda,m\right\rangle\right\|^2=  \left\langle\lambda,m \right| X_\pm X_\mp \left|\lambda,m\right\rangle
    =m(m\mp 1) -\lambda \geq 0\,.
\end{equation}

Therefore,  $\left(m\mp \frac{1}{2}\right)^2\geq \lambda+\frac{1}{4}$. Two cases should be considered \cite{Bargmann}: $\lambda+\frac{1}{4} \geq 0$ and $\lambda+\frac{1}{4} < 0$, which give rise to the so-called \emph{discrete} and \emph{continuous} classes of unitary \emph{irrep}'s respectively.

\subsubsection{Discrete classes: $\lambda+\frac{1}{4} \geq 0$}\label{discretas}

Let us write $\lambda= k(k-1)$ with\footnote{The case $0\leq k < \frac{1}{2}$ can be mapped onto the one considered through the change $k'=1-k>\frac{1}{2}$, since $k'(k'-1)=k(k-1)$.} $k \geq \frac{1}{2}$. Then, $\lambda + \frac{1}{4}= \left( k-\frac{1}{2} \right)^2\geq 0$. Then, we have  either $m\geq k$ or $m\leq -k$.

From (\ref{20}) it follows that the existence of a vector with, respectively, a minimum or a maximum eigenvalue $m_0$ requires that
\begin{equation}\label{22}
    m_0(m_0\mp1)-k(k-1)=(m_0\mp k)\left(m_0\pm(k-1)\right)=0 \quad \Rightarrow \quad m_0=\pm k\ {\rm respectively.}
\end{equation}
Therefore, in these \emph{irrep}'s also $k$  takes integer or half-integer values, $k=\frac{1}{2}, 1, \frac{3}{2}, \cdots$ (which justifies the name of discrete assigned to these two classes).

The subsequent application on these vectors of $X_+$ and $X_-$ respectively generates an infinite sequence of eigenvectors of $X_0$ corresponding to the eigenvalues $m = k +n$ or $m=-k-n$ respectively, with $n \in \mathbb{N} \cup \left\{ 0 \right\}$.

\medskip

These \emph{irrep}'s can be explicitly  realized on a space of analytic functions of a complex variable which are regular on the open unit circle \cite{Bargmann,Jackiw}. Indeed, let us consider the Hilbert space defined by the set of functions $f(\mathfrak{z})$ analytic on the open disk $\mathcal{M}:=\left\{\mathfrak{z}\in \mathbb{C} : |\mathfrak{z}|<1\right\}$ with the scalar product
\begin{equation}\label{ana1}
    \begin{array}{c} \displaystyle
      \left(f(\mathfrak{z}),g(\mathfrak{z})\right)_k:=\frac{2k-1}{\pi} \int_{\mathcal{M}} \frac{d\mathfrak{z}\, d\bar{\mathfrak{z}}}{2 \imath} \left[1-\bar{\mathfrak{z}} \mathfrak{z} \right]^{2(k-1)} {f(\mathfrak{z})}^* g(\mathfrak{z})
      \\ \\ \displaystyle
      =  \frac{2k-1}{2\pi} \int_{0}^{2\pi} d\phi \int_0^{1}d r^2 \left[1-r^2 \right]^{2(k-1)} {f(r e^{\imath \phi})}^* g(r e^{\imath \phi})\,,
    \end{array}
\end{equation}
with $k>\frac{1}{2}$. This definition can be extended to $k=\frac{1}{2}$ as \cite{Bargmann}
\begin{equation}\label{ana2}
    \left(f(\mathfrak{z}),g(\mathfrak{z})\right)_{1/2}:=\lim_{k\rightarrow {\frac{1}{2}}^+}\frac{2k-1}{\pi} \int_{\mathcal{M}} \frac{d\mathfrak{z}\, d\bar{\mathfrak{z}}}{2 \imath} \left[1-\bar{\mathfrak{z}} \mathfrak{z} \right]^{2(k-1)} {f(\mathfrak{z})}^* g(\mathfrak{z})\,.
\end{equation}

An orthonormal and complete basis of this space can be constructed as
\begin{equation}\label{ana3}
    \left\{ h_l(\mathfrak{z}):=\left(\frac{\Gamma(2k+l)}{\Gamma(2k)\Gamma(l+1)} \right)^{\frac{1}{2}} \mathfrak{z}^l\,, l=0,1,2,\dots\right\}
\end{equation}
and it can be shown that, for any square-integrable function in this space, the series $f(\mathfrak{z})=\sum_{l=0}^\infty c_l h_l(\mathfrak{z})$ also converges in a pointwise sense and $f(\mathfrak{z})$ is regular on the open disk $\mathcal{M}$ \cite{Bargmann}.

It can be straightforwardly verified that the differential operators \cite{Bargmann}
\begin{equation}\label{ana4}
    X_0:=\mathfrak{z} \partial_\mathfrak{z} +k\,, \quad X_+:=-\mathfrak{z}^2 \partial_\mathfrak{z} -2k \mathfrak{z} \,, \quad X_-:=-\partial_\mathfrak{z}
\end{equation}
are a realization of the generators of $sl(2,\mathbb{R})$ in Eq.\ (\ref{17}) and their Hermitian conjugates in this space satisfy ${X_0}^\dagger=X_0$ and ${X_{\pm}}^\dagger=X_{\mp}$.

Moreover,
\begin{equation}\label{ana5}
    X_0 h_l(\mathfrak{z})=m  h_l(\mathfrak{z})\quad {\rm with}\quad m=l+k\,, l=0,1,2,\dots
\end{equation}
and
\begin{equation}\label{ana6}
    \mathbf{X}^2 =  \lambda \mathbf{1}\,, \quad {\rm with}\quad \lambda=k(k-1)\,,
\end{equation}
which corresponds to a unitary \emph{irrep} with a minimal eigenvalue for $X_0$, $m_0=k$, with $k=\frac{1}{2},1,\frac{3}{2},\dots$

A unitary \emph{irrep} with a maximal eigenvalue for the rotations generator is obtained by taking on the same space \cite{Bargmann}
\begin{equation}\label{ana7}
    X'_0:=-X_0=-\mathfrak{z} \partial_\mathfrak{z} - k\,, \quad X'_+:=-X_-=\partial_\mathfrak{z}\,, \quad X'_-:= -X_+=\mathfrak{z}^2 \partial_\mathfrak{z} +2k \mathfrak{z} \,,
\end{equation}
with $\lambda=k(k-1)$ and $m=-k,-k-1,-k-2,\dots$

\subsubsection{Continuous classes: $\lambda+\frac{1}{4} < 0$}\label{continuas}

In this case we write $\lambda=k(k-1)$ with $k=\frac{1}{2}+\imath \gamma$ and $\gamma\in\mathbb{R}$. Then, $\lambda+\frac{1}{4}=-\gamma^2<0$ and the condition in Eq.\ (\ref{20}) reduces to  $\left(m\mp \frac{1}{2}\right)^2\geq 0 >-\gamma^2$, satisfied for any integer or half-integer value of $m$. We take $\gamma>0$, which justifies the name of \emph{continuous} given to these classes of \emph{irrep}'s.

In such a way, $m$ is not bounded and takes either all the integer or all the half-integer values. Moreover, in these \emph{irrep}'s the Casimir invariant takes only negative values, $\mathbf{X}^2=-\left(\gamma^2+\frac{1}{4}\right) \mathbf{1}$.

The unitary representations corresponding to these two continuous classes can be explicitly realized as a function space over the unit circle, as discussed in \cite{Bargmann}. Indeed, let us consider the Hilbert space of function $f(\phi)$ defined on the closed interval $[0,2\pi]$ with the scalar product defined with the usual Lebesgue measure
\begin{equation}\label{circ1}
    \left( f(\phi),g(\phi) \right):=\int_0^{2\pi}d\phi\, {f(\phi)}^* g(\phi)\,.
\end{equation}

On this space we define \cite{Bargmann}
\begin{equation}\label{circ2}
      X_0:= -\imath \partial_\phi\,, \quad
      X_+:=e^{\imath \phi}\left(\imath \partial_\phi -\frac{1}{2}-\imath \gamma \right)\,, \quad
      X_-:= e^{-\imath \phi}\left(\imath \partial_\phi +\frac{1}{2}+\imath \gamma \right)\,,
\end{equation}
with real $\gamma$. It is a straightforward exercise to verify that these operators satisfy the commutation relations in Eq.\ (\ref{17}) and its Hermitian conjugate are given by ${X_0}^\dagger={X_0}$ and ${X_\pm}^\dagger=X_\mp$ if defined on a domain of periodic or antiperiodic functions on the interval $[0,2\pi]$. Moreover,
\begin{equation}\label{circ3}
    \mathbf{X}^2=-\left(\frac{1}{4}+\gamma^2\right) \mathbf{1}\,.
\end{equation}

Therefore, for any $\gamma>0$ and adopting periodic boundary conditions, $f(2\pi)=f(0)$, we can take the complete orthonormal basis
\begin{equation}\label{circ4}
    \left\{ h_m(\phi):= \frac{1}{\sqrt{2\pi}}\,e^{\imath m \phi}\,, m\in \mathbb{Z}  \right\}
\end{equation}
and have for these vectors
\begin{equation}\label{circ5}
    X_0 h_m(\phi)=m h_m(\phi)\,, \quad {\rm with} \quad m=0,\pm 1, \pm 2  , \dots
\end{equation}

On the other hand, if we adopt antiperiodic boundary conditions, $f(2\pi)=-f(0)$, we can take the complete orthonormal basis
\begin{equation}\label{circ6}
    \left\{ h'_m(\phi):= \frac{1}{\sqrt{2\pi}}\,e^{\imath m \phi}\,, m\in \mathbb{Z}+\frac{1}{2}  \right\}
\end{equation}
having  for these vectors
\begin{equation}\label{circ7}
    X_0 h'_m(\phi)=m h'_m(\phi)\,, \quad {\rm with} \quad m=\pm \frac{1}{2}, \pm \frac{3}{2} , \pm \frac{5}{2},  \dots
\end{equation}

%%%%%%%%%%%%%%%%%%%%%%%%%%%%%%%%%%%%%%%%%%%%%%%%%%%%%

%%%%%%%%%%%%%%%%%%%%%%%%%%%%%%%%%%%%%%%%%%%%%%%%%%%%%

\end{document}